\documentstyle{mn}
\input epsf

\newcommand{\msigma}{$M_{\bullet}$--$\sigma_*$\ }
\newcommand{\msun}{M_{\odot }}

\newcommand{\kms}{\, {\rm km \, s}^{-1} }

\newcommand{\yr}{{\, \rm yr} }

\title{Multiple black holes in galactic bulges}

\author[Haehnelt and Kauffmann]
       {Martin G. Haehnelt and Guinevere Kauffmann\\
        Institute of Astronomy, Madingley Road, Cambridge CB3 0HA\\ 
        Max-Planck-Institut f\"ur Astrophysik,
        Karl-Schwarzschild-Str. 1, D-85748 Garching}

\pagerange{\pageref{firstpage}--\pageref{lastpage}}
\begin{document}

\maketitle

\label{firstpage}

\begin{abstract}
We study the number and interaction rates of supermassive black holes in 
galactic bulges as predicted by hierarchical models of galaxy formation in which
the spheroidal components of galaxies are formed by mergers.
In bright ellipticals, the number of events that can eject a central supermassive binary black
hole is large.  Central binaries     
must  therefore merge in less than a Hubble time -- otherwise 
there will be too much  scatter in the \msigma relation and 
too many off-center galactic nuclei. We propose 
that binary black holes are able to merge during the major gas  accretion
events that trigger QSO activity in galaxies. If this is the case,           
less than  10 percent of faint ellipticals and 
40 percent of bright ellipticals are predicted to harbour binary black holes
with near equal masses at their centres. This binary may be ejected away
from the centre of the galaxy or even into intergalactic space 
in up to 20\% of the  most luminous ellipticals.
The number of low mass black holes that can interact with the central object is predicted
to be a strong function of galaxy luminosity. In most faint ellipticals, 
no black holes fall into the centre of the galaxy after the last major gas accretion event,
but in the most luminous ellipticals, an average of 10 low mass black holes
interact with the central supermassive object after this time.         
It is expected that stars will be ejected from galaxy cores as these
low mass-ratio binaries harden. Multiple black holes in galactic bulges thus  provide
a natural explanation for the strong systematic trends in the
observed central density profiles of ellipticals as a function of luminosity. 
\end{abstract}


\section{Introduction}

According to the standard paradigm of structure formation in the Universe,
galaxies merge frequently as their dark matter halos assemble. 
Frequent galaxy mergers              
will inevitably lead to the  the formation of supermassive binary 
black holes (e.g. Miloslajevic \& Merritt 2001).  
It is not clear if these supermassive binaries will be able to  merge  
on a  Hubble time (Begelman, Blandford \& Rees 1980; 
Milosavljevic \& Merritt 2001; Yu 2002).  
If  merging timescales are long,  the binary is likely to interact  
with new infalling black holes through  gravitational 
slingshot interactions (Saslaw, Valtonen \& Aarseth  1974). There is some
circumstantial observational evidence that binary black holes do merge
(e.g.  Meritt \& Ekers 2002) , but very little is known about the
rate at which these mergers occur.

A number of authors have presented models for the growth  of 
supermassive black holes in  merging galaxies in a cold dark matter (CDM) Universe  
(Cattaneo, Haehnelt\& Rees 1999;  Kauffmann \& Haehnelt 2000 (KH2000);  
Monaco, Salucci \& Danese 2000; Menou,  Haiman \& Narayanan 2001; 
Hein\"am\"aki 2001; Volonteri, Haardt \& Madau 2002, Cattaneo 2002). 
Here, we extend the models of KH2000, which  explicitly
followed the star formation  histories and dynamical evolution of bulges. 
In a subsequent paper, Haehnelt \& Kauffmann (2000)
demonstrated that the same model could also reproduce the observed 
\msigma relation (Gebhardt et al. 2000, Ferrarese \& Merrit 2000). 
KH2000 made the extreme assumption that  black holes merge
instantaneously  when a bulge forms through a merger of two galaxies of roughly equal mass.
In this Letter we relax this assumption 
and we discuss some of the issues which will influence our predictions of the
multiplicity of black
holes  in galactic bulges. We also explore the possibility that 
the merging of supermassive binary black holes in bright ellipticals 
is responsible for the observed cusp/core  dichotomy of elliptical
galaxies (Gebhardt et al. 1996; Merritt 2001).   

\section{The build-up of supermassive black holes in KH2000}

In the model of KH2000, spheroids form when two galaxies of comparable
mass merge.  
The central black holes of the progenitor galaxies     
are assumed to coalesce  instantaneously and a fraction of the 
available cold gas is accreted.  The amount of  cold 
gas in galaxies is determined by the balance of cooling, star formation and ``feedback'' by supernovae.  
The fraction of the available cold gas accreted by the black hole is assumed to scale
with the circular velocity $v_{\rm circ}$  of the surrounding dark
matter halo. Note that in the models, the mass accretion rate often exceeds the
Eddington rate at high redshifts.

Figure 1 shows a typical  accretion 
and merging history of the central supermassive black hole of a 
bright  elliptical galaxy  (present-day luminosity $M_V=-22.9$). 
In the KH2000 models,  the black holes typically
``form'' at $5<z_{\rm form}<10$ in potential wells of
$v_{\rm c} \ga 100 \kms$  by direct  collapse of the available cold gas 
to the centre of the two merging galaxies.  
In the case of the galaxy  in Fig. 1 a black hole of 
$3\times 10^{6} \msun$ forms at $z=8$ in this way. 
Initially the black hole mass grows during a few events    
in which the accreted gas mass exceeds the total mass of the two 
black holes in the progenitor galaxies. 
At low redshifts, bright ellipticals  contain little gas
and the black hole mass grows more slowly  by merging of additional black holes
and not by the infall of large quantities of new cold gas.
Note that in the model of KH2000, the amount of cold gas available for fueling 
black holes decreases rapidly with decreasing  circular velocity below 
$v_{\rm c}\sim 200 \kms$, because of the increasing impact of  
supernovae feedback in shallow potential wells. As a result, few                        
black holes with masses smaller than a few times $10^{5} \msun$ form.
In the example presented here, a  total of  about 30 black holes have 
fallen into the galaxy by the present day.

\begin{figure}
\centerline{
\epsfxsize=8cm \epsfbox{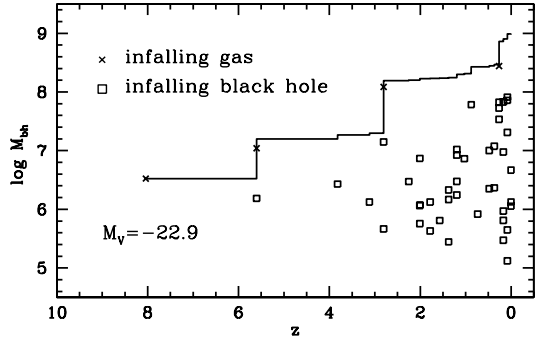}
}
\caption{\label{fig1}
\small Typical accretion/merging history of the central black hole 
in a bright elliptical galaxy ($M_V = -22.9$). Crosses denote 
the gas infall during phases of QSO activity. Squares show the 
masses of infalling black holes. 
}                  
\end {figure}
\normalsize

\section{Multiple black holes} 

\subsection{Formation and merging of binary  black holes}

When two galaxies merge, the smaller galaxy will                              
sink to the centre of the merger remnant  because of dynamical
friction. The outer regions of the infalling galaxy 
will be gradually tidally stripped in the process. 
If two galaxies with roughly equal mass merge, 
a binary black hole will form within a few dynamical times 
 (Miloslajevic \& Merrit 2001). If the mass ratio between the two galaxies is large,
the time it takes for the infalling galaxy to reach the center of the remnant
also becomes large and may exceed the Hubble time.
In practice, the timescale for the formation of the binary will depend 
not only on the mass ratio of the galaxies, but also on the 
density profiles of the merging galaxies and the orbital 
parameters of the infalling galaxy.

The subsequent evolution of the supermassive binary has been discussed 
by Begelman, Blandford \& Rees (1980). 
 For an initially circular orbit 
coalescence within time T occurs when the velocity of the orbit $v_{\rm orb}$ 
reaches
\begin{eqnarray} 
v_{\rm orb} \sim v_{\rm gr} &\sim& 2600   
\left (\frac{m_{\rm prim}+m_{\rm sec}}{10^8\msun}\right )^{1/8}
\left ( \frac{m_{\rm sec}}{m_{\rm prim}}\right)^{-1/8} \nonumber \\
&&\left ( \frac{T}{10^{10} \yr}\right)^{-1/8}  \kms ,  \\
\nonumber
\end{eqnarray} 
where $m_{\rm prim}$ and $m_{\rm sec}$ are the mass of the primary and
secondary black holes 
(Peters 1964, Gould \& Rix 2000).
It is uncertain whether stellar-dynamical  processes will be able to reduce the 
separation of the binary so that
coalescence by emission of gravitational  waves 
will occur in less than a Hubble time.                                                      

The binary is expected to harden either by gravitational sling-shot ejection of
stars (Quinlan 1996, Milosvlajevic \& Meritt 2001) 
or by the accretion of gas onto the binary system (Armitage \& Natarajan 2002). 
The timescale for the ejection of stars to cause the binary to harden 
will exceed the Hubble time in bright galaxies unless stars are scattered 
into low-angular momentum orbits  more efficiently than 
by star-star relaxation (see Zhao, Haehnelt \& Rees 2002 for some
suggestions how that may be achieved) 
or the central regions of galaxies are sufficientlty triaxial 
(Begelman, Blandford, \& Rees 1980, Yu 2002).

The accretion of gas can in principle reduce the binary 
separation  on a much shorter timescale if the accreted gas mass 
exceeds that of the binary but it is  uncertain how efficiently this process operates. 

\begin{figure}
\centerline{
\epsfxsize=8cm \epsfbox{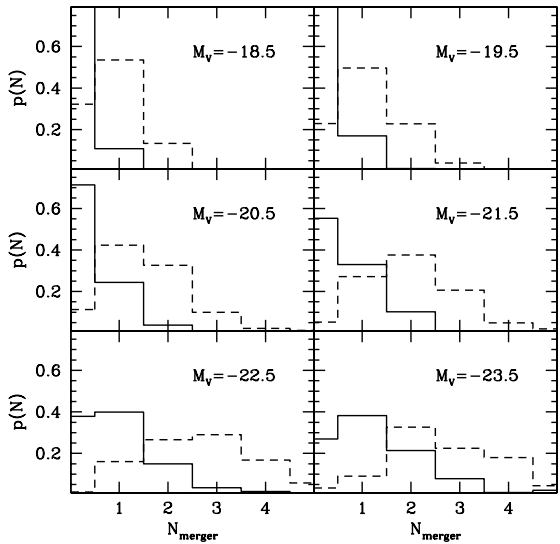}
}
\caption{\label{fig2}
\small The distribution function of the number of black hole mergers with mass ratios 
$m_{\rm sec}/m_{\rm prim}>0.3$. Results are shown for galactic bulges with different 
V-band luminosities. The dashed histogram shows  the total number of
possible black hole mergers. 
The solid histogram is the number of mergers 
{\it after} the last major  gas accretion event, in which the  accreted gas 
mass exceeds the sum of the masses of the two black holes 
in the merging galaxies.  
}                  
\end {figure}
\normalsize

\subsection{Mergers of equal mass black holes and the expected binary fraction}

After a hard binary has formed 
with a  circular velocity larger than the stellar velocity dispersion of
the galaxy, 
the infall of a third black hole will normally  lead to 
gravitational slingshot ejection of the lightest black hole 
with a  velocity of around one third of the relative orbital velocity of the
binary black hole (Saslaw et al. 1974, Hut and Rees 1992). 
The binary will also  get a kick velocity  
that is smaller than the velocity of the ejected black hole by a factor $ m_{\rm ej}/ m_{\rm bin}$. 
The time evolution of the hardening is uncertain and it is
difficult to predict what the orbital velocity of the supermassive
binary is likely to be for a typical slingshot ejection. Initially, when low
angular momentum stellar orbits are still populated, the binary will 
harden quickly.  If all three black holes have similar masses,  
the kick velocity will be sufficient to kick the binary 
into the outer parts of the galaxy or even to eject it entirely
(see Hut \& Rees 1992 for a discussion).     

The dashed line in Figure 2 shows the distribution of  
the number of black hole  mergers in elliptical galaxies  
with  mass ratios $m_{\rm sec}/m_{\rm prim}>0.3$. Results are shown  for 
a range  of bulge luminosities.  The median number of mergers
increases from one in faint bulges to three in bright bulges.  
Figure 2 demonstrates that if the merging timescale 
of supermassive binary black holes is longer than                
the Hubble time,  a binary should be ejected in up to 40 percent 
of bright elliptical galaxies.
This would appear to conflict with the fact that black holes are 
observed in all
nearby bright elliptical galaxies  and with the  tightness of the
observed \msigma relation. 
This strongly supports the hypothesis that the merging timescale of
supermassive binaries is shorter than the Hubble time   
(Zhao et al. 2002, Merritt \& Ekers 2002).  

The solid line in Figure 2 shows the distribution of the  number 
of mergers  after  the last accretion event in which the mass     
of accreted gas exceeded the sum of the masses of the two black holes 
in the binary system. 
The median number of mergers since this event ranges from zero 
in faint galaxies to one in bright galaxies. 
If the supermassive  black holes do indeed merge during gas-rich
accretion events, the  
fraction of elliptical galaxies containing large mass ratio binary black holes 
will not be larger than 10\%
in faint ellipticals and 40\% in brighter objects.
The fraction of galaxies with  a third massive ``intruder''   
ranges  from 0 to 20 percent. Binary black hole 
ejections will then only occur in a small fraction  
of only  the  brightest  galaxies. 

\subsection{Unequal mass mergers and the                           
core/cusp dichotomy of elliptical galaxies}

Observed bulges  show a clear dichotomy in their central 
density profiles.  Faint galaxies have steep inner density 
profile which are close to isothermal ($\rho \propto r^{-\gamma}$,
$\gamma=2$) while bright galaxies have significantly shallower 
core profiles ($\gamma \la 1$). There is  a transition from cusp to core profiles
at a V-band luminosity of around -20  
 (Gebhardt et al. 1996).  Some bright galaxies even show 
indications of a  decreasing stellar density at their centres 
(Lauer et al. 2002). 

It has been argued that the growth of
supermassive black holes may affect the central stellar density
profile of galactic bulges beyond the radius                                 
where the black hole dominates the gravitational potential. 
Even the sign of the effect remains controversial, however.  
Adiabatic growth of a single black hole would lead to a 
steepening of the density profile  (van der Marel 1999), 
while  binary hardening due to stellar dynamical interactions will deplete the 
central regions and lead to 
a flattening of the density profile (e.g. Merritt 2001, Miloslavjevic \& Meritt 2001). 
Quinlan (1996) studied the  hardening 
of supermassive binary black holes
and found that small mass ratio binaries eject stars nearly 
as efficiently as large mass ratio binaries. 
This suggests that the total number of mergers may determine  the 
ejected mass.
Ravindranath et al (2002) and Milosavljevic et al. (2002) assumed that
the density profiles of elliptical galaxies are always initially steep
and calculated the mass in stars that would have to be ejected from the
centre in order to explain the observed profiles of bright ellipticals.
They found a strong non-linear correlation between the observed black hole mass
and the ejected mass. The ejected mass is larger than the black hole mass by
a factor of a few. 

\begin{figure}
\centerline{
\epsfxsize=8cm \epsfbox{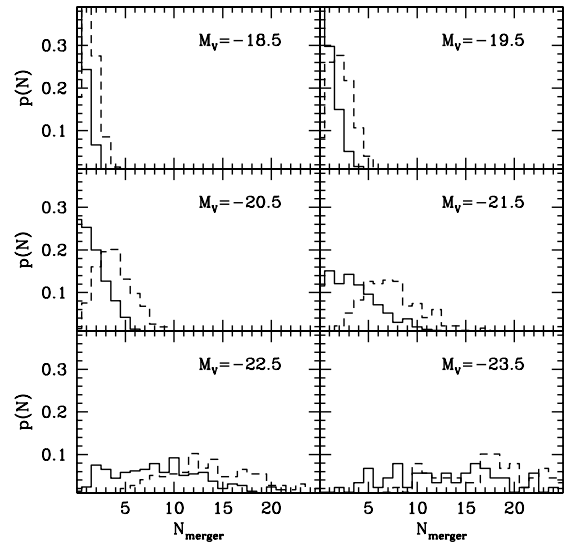}
}
\caption{\label{fig3}
\small
As in figure 2, except for black hole mergers with  mass ratios
$m_{\rm sec}/m_{prim}>0.01$. The mass ratio of the two merging
galaxies has also been constrained to be larger than 0.01.
Dashed histogram is the total number of mergers 
in the model. The solid histogram is the number of mergers 
{\it after} the last major gas-rich accretion event.}  
\end {figure}
\normalsize

It is not understood, however, why there is a 
dichotomy in profile shape between bright and faint galaxies. It is also
not clear whether black holes merge
frequently enough to eject a few times the mass of the 
binary.  
In Figure 3 we show  the number of black hole  mergers
with  mass ratios  $m_{\rm sec}/m_{\rm prim}>0.01$
that occur after the last major gas accretion event. 
Van den Bosch et al. (1999) studied  the sinking of DM satellites within
DM haloes and found that 
for orbits with moderate ellipticities (0.3-0.5) the timescale to reach the centre 
approaches the Hubble time for mass ratios smaller than $0.01$.  
The situation may not be exactly analogous for mergers of galaxies
rather than dark matter halos, because the density profiles
of galaxies are different and they also contain gas. 
Nevertheless, in order to exclude black holes that may take more than a Hubble time
to reach the centre of the parent galaxy, we  only count mergers with  $m_{\rm sat}/m_{\rm prim}>0.01$. 
The median number of such minor mergers increases from 0 for faint ellipticals to 10
for the brightest systems.
More than 50 \% of  galaxies fainter than $M_V =-20$ do not experience any black hole
mergers,  while  the mean/median number of 
mergers rises strongly for galaxies brighter than $M_V =-20$ (Fig.4). 

The relation  between the total number of mergers and the actual ejected mass 
is uncertain. In order to predict the ejected mass, we would need to
know the                            
relaxation time scale on which low angular momentum orbits 
are repopulated and the detailed orbital structure 
of the central part of elliptical galaxies. Nevertheless  our results show that
there is a strong 
dependence of the total number of mergers on galaxy luminosity 
and this may well explain the core/cusp dichotomy of elliptical
galaxies. 
The large ejected masses in bright 
galaxies would require efficient hardening in  5 to 10
merging events.



\begin{figure}
\centerline{
\epsfxsize=8cm \epsfbox{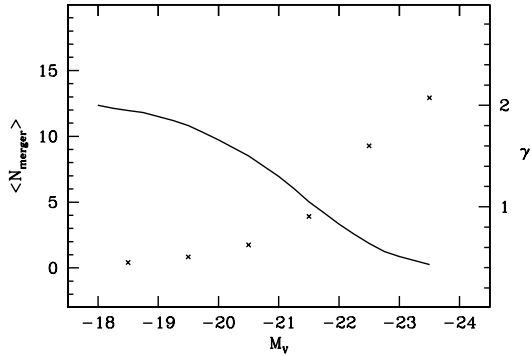}
}
\caption{\label{fig3}
\small
The crosses show the mean  number of black hole mergers with mass ratios 
$m_{\rm sec}/m_{\rm prim}>0.01$  occurring after the
last gas-rich accretion event,  as a function of the V-band magnitude of the bulge.
The solid curve shows the observed power law index $\gamma$ of the 
central density profile of galactic bulges as a function of luminosity 
(Gebhardt et al. 1996, Merritt 2001). 
}                  
\end {figure}
\normalsize

\section{Conclusions}

According to hierarchical merger models, the formation of supermassive 
binary black holes will be common. Unless 
the majority of these binaries merge faster than a Hubble time, 
further infall of new black holes will lead                           
to an ejection rate of these  binary systems that    
is too large  to be consistent with the small scatter 
in the \msigma relation. There is also no observational evidence 
for the existence of a significant fraction of galaxies with off-centre galactic nuclei. 
The accretion of gas  
and binary hardening by stars are the two main candidates 
for driving the merger of these binaries.  We have assumed  that 
the former is efficient if the accreted gas mass exceeds the total mass 
of the supermassive binary. This reduces the  fraction
of faint ellipticals with binary black holes to less than 10\% 
and the fraction of bright galaxies that harbour binaries to
less than 40\%.
Up to  20\% of the brightest
ellipticals have  recently ejected a binary from their centres.  
We  thus predict that a small fraction  of  bright ellipticals do not contain
a central black hole with a mass that fits onto the 
\msigma relation. We note, however, that the fraction of binaries that
are truly ejected from the galaxy depends sensitively on the   time evolution  
of the hardening and on the detailed orbital structure 
at the centre of galactic bulges.  
Gas accretion in cooling flows may also prevent efficient binary ejection. 

Finally, we have demonstrated that the total number of expected low
mass ratio black hole mergers 
is a strong function 
of bulge luminosity. A strong transition from no mergers to many mergers
occurs at the  same characteristic
luminosity ($M_V =-20$) where ellipticals transition from cuspy to
core-dominated central profiles.
This supports the hypothesis that the ejection of stars 
by central supermassive binary black holes is indeed responsible for the shallow
central density profiles observed in luminous elliptical galaxies.


\label{lastpage}

\end{document}